\documentclass{ws-procs9x6}

\usepackage{xspace}

\newlength{\hoehe}

\def\Cite#1{\protect\citelow{#1}}

\def\cref#1{Chapt.\,\ref{#1}}
\def\Cref#1{Chapter~\ref{#1}}
\def\sref#1{Sect.\,\ref{#1}}

\def\fref#1{Fig.\,\ref{#1}}

\def\eref#1{(\ref{#1})}
\def\Eref#1{Equation (\ref{#1})}

\def\rref#1{Ref.\,\citelow{#1}}

\def\lleft{\textit{left}}
\def\rright{\textit{right}}
\def\LLeft{\textit{Left}}
\def\RRight{\textit{Right}}

\def\1{\footnotemark[1]}
\def\and{\& }
\def\Cerenkov{\v{C}erenkov\xspace}

\def\gcm2{g/cm$^2$\xspace}
\def\li{$\lambda_i$\xspace}
\def\lg{{\rm lg}\xspace}

\def\Xmax{$X_{max}$\xspace}

\def\lnA{\langle\ln A\rangle}

\def\line{---}
\def\dashed{-\,-\,-}

\begin{document}

\title{COSMIC RAYS FROM THE KNEE TO THE SECOND KNEE: $10^{14}$ TO $10^{18}$~eV 
 \footnote{Lecture given at the International School of Cosmic Ray
           Astrophysics, 15th Course: Astrophysics at Ultra-high Energies,
           20-27 June 2006, Ettore Majorana Centre Erice, Sicily, Italy}}

\author{J\"ORG R.~H\"ORANDEL}

\address{University of Karlsruhe, Institute for Experimental Nuclear Physics,
         P.O. 3640\\ 76021 Karlsruhe, Germany --- www-ik.fzk.de/$\sim$joerg}

\begin{abstract}
 The energies of cosmic rays, fully ionized charged nuclei, extend over a wide
 range up to $10^{20}$~eV. A particularly interesting energy region spans from
 $10^{14}$ to $10^{18}$~eV, where the all-particle energy spectrum exhibits two
 interesting structures, the 'knee' and the 'second knee'. An explanation of
 these features is thought to be an important step in understanding of the
 origin of the high-energy particles.  Recent results of air shower experiments
 in this region are discussed.  Special attention is drawn to explain the
 principle of air shower measurements --- a simple Heitler model of (hadronic)
 air showers is developed.
\end{abstract}

\keywords{cosmic rays, knee, air showers, Heitler model}

\bodymatter

\section{Introduction}

\begin{figure}[t]\centering
  \psfig{file=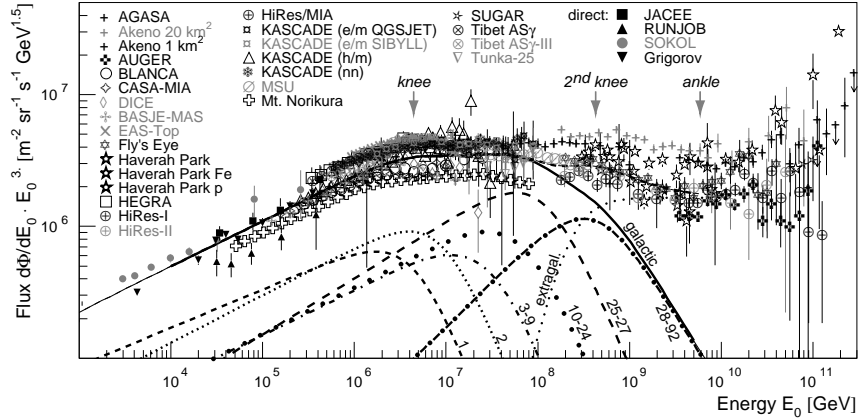,width=0.99\textwidth}%
  \caption{All-particle energy spectrum of cosmic rays, the flux is multiplied
	   by $E^3$, for references see\Cite{aspenreview}. The lines represent
	   spectra for elemental groups (with nuclear charge numbers $Z$ as
	   indicated) according to the poly-gonato model\Cite{pg}. The sum
	   of all elements (galactic) and a presumably extragalactic component
	   are shown as well.}
  \label{espec}	   
\end{figure}

The energy spectrum of cosmic rays (fully ionized atomic nuclei) spans a wide
range in energy from GeV energies up to $10^{20}$~eV. Over these 10 decades the
flux decreases by about 30 orders of magnitude rather featureless, following
roughly a power law $dN/dE\propto E^\gamma$. The power law behavior indicates a
non-thermal origin of the particles. To reveal small structures in the shape of
the energy spectrum the flux is usually multiplied by the energy to some power.
The energy spectrum multiplied by $E^3$ is depicted in \fref{espec}. In this
representation the spectrum looks rather flat and fine structures can be
recognized, indicating small changes in the spectral index $\gamma$. The most
important are the {\sl knee} at $E_k\approx4.5$~PeV where the power law
spectral index changes from $\gamma=-2.7$ at low energies to
$\gamma\approx-3.1$, the {\sl second knee} at
$E_{2nd}\approx400$~PeV$\approx92\times E_k$, where the spectrum exhibits a
second steepening to $\gamma\approx-3.3$, and the {\sl ankle} at about 4~EeV,
above this energy the spectrum seems to flatten again to $\gamma\approx-2.7$.
To understand the origin of these structures is expected to be a key element in
the understanding of the origin of cosmic rays (CRs).

The lecture starts with a short overview on the physics of galactic cosmic rays
(\sref{physs}). Measurements in the energy region of interest are performed
with air shower experiments, their principles are outlined in \sref{eass} with
a simple Heitler model.  Finally, recent results of air shower experiments are
reviewed in \sref{ress}.

\section{Galactic Cosmic Rays and the Knee} \label{physs}
\subsection{Sources}
At energies around 1~GeV/n all elements known from the periodic table with
nuclear charge number $Z$ from 1 to 92 have been found in
CRs\citelow{shapirosilberberg,wiebel,pg}. Overall, the abundance of elements in
CRs is very similar to the abundance found in the solar system, which indicates
that CRs are "regular matter" but accelerated to very high energies.  This is
emphasized by measurements of the CRIS experiment \citelow{cris-abundance}
which show that the abundances of particular isotopes in cosmic rays and in the
solar system differ by less than 20\%.

The bulk of CRs is assumed to be accelerated in blast waves of supernova
remnants (SNRs). This goes back to an idea by Baade and Zwicky who proposed
SNRs as cosmic-rays sources due to energy balance
considerations\citelow{baadezwicky}. They realized that the power necessary to
sustain the cosmic-ray flux could be provided when a small fraction $\sim10\%$
of the kinetical energy released in supernova explosions is converted into CRs.
Fermi proposed a mechanism to accelerate particles with moving magnetic
clouds\citelow{fermi}. This led to todays picture that the particles are
accelerated at strong shock fronts in SNRs through first-order Fermi
acceleration\citelow{axford,krymsky,bell,blanford,longair2}. This theory predicts
spectra at the sources following a power law $dN/dE\propto E^{-2.1}$.

Diffusive, first-order shock acceleration works by virtue of the fact that
particles gain an amount of energy $\Delta E\propto E$ at each cycle, when a
cycle consists of a particle passing from the upstream (unshocked) region to
the downstream region and back.  At each cycle, there is a probability that the
particle is lost downstream and does not return to the shock. Higher energy
particles are those that remain longer in the vicinity of the shock and have
enough time to achieve the high energy. After a time $T$ the maximum energy
attained is $E_{max}\sim Z e \beta_s B T V_s$, where $\beta_s=V_s/c$ is the
velocity of the shock.  This results in an upper limit, assuming a minimal
diffusion length equal to the Larmor radius of a particle of charge $Ze$ in the
magnetic fields $B$ behind and ahead of the shock.  Using typical values of
Type II supernovae exploding in an average interstellar medium yields
$E_{max}\approx Z \cdot100$~TeV\citelow{lagagemax}.  More recent estimates give a
maximum energy up to one order of magnitude larger for some types of supernovae
$E_{max}\approx Z\cdot5$~PeV\citelow{berezhkomax,kobayakawa,sveshnikova}. As the
maximum energy depends on the charge $Z$, heavier nuclei (with larger $Z$) can
be accelerated to higher energies. This leads to consecutive cut-offs of the
energy spectra for individual elements proportional to their charge $Z$,
starting with the proton component.

This theory is strongly supported by recent measurements of the HESS
experiment\citelow{hesssnr,hessrxj1713}, observing TeV $\gamma$-rays from the
shell type SNR RX J1713.7-3946. For the first time, a SNR could be spatially
resolved in $\gamma$-rays and spectra have been derived directly at a potential
cosmic-ray source. The measurements yield a spectral index
$\gamma=-2.19\pm0.09\pm0.15$ for the observed $\gamma$-ray flux. The results
are compatible with a nonlinear kinetic theory of cosmic-ray acceleration in
supernova remnants and imply that this supernova remnant is an effective source
of nuclear CRs, where about 10\% of the mechanical explosion energy are
converted into nuclear CRs\citelow{voelkrxj1713}.

\subsection{Propagation}
After acceleration, the particles propagate in a diffusive process through the
Galaxy, being deflected many times by the randomly oriented magnetic fields
($B\approx3$~$\mu$G). The nuclei are not confined to the galactic disc, they
propagate in the galactic halo as well. The diffuse $\gamma$-ray background,
extending well above the disc, detected by the EGRET experiment, exhibits a
structure in the GeV region, which is interpreted as indication for the
interaction of propagating CRs with interstellar
matter\citelow{strong-moskalenko}.  The $\gamma$-rays are produced in inelastic
hadronic reactions of CRs with the interstellar medium (ISM) via neutral pion
decay $p+\mbox{ISM}\rightarrow\pi^0\rightarrow\gamma\gamma$.  The height of the
propagation region in the halo has been estimated measuring the
$^{10}$Be/$^9$Be-ratio with the ISOMAX experiment to be a few kpc
\citelow{simon-height}.  Determining the abundance of radioactive nuclei, which
decay on the way from the source to the Earth, allows to determine the
residence time of CRs in the Galaxy. Measurements with the CRIS instrument
yield about $15\cdot10^6$~a for particles with GeV energies\citelow{cris-time}.

Information on the propagation pathlength of CRs is often derived from the
measurement of the ratio of primary to secondary nuclei. The latter are
produced through spallation during propagation in the Galaxy.  The energy
dependence of the measured ratio is frequently explained in Leaky Box models by
a decreasing pathlength of CRs in the Galaxy $\Lambda(R) = \Lambda_0
(R/R_0)^{-\delta}$, with typical values $\Lambda_0\approx10 - 15$~g/cm$^2$,
$\delta\approx0.5 - 0.6$, and the rigidity $R_0\approx4$~GV\citelow{stephens}.
In this picture the spectra observed at Earth should be steeper as compared to
the source, i.e.\ the spectral index $\gamma$ should be smaller by the value of
$\delta$.

Energy spectra of individual elements have been measured up to energies of
about $10^{14}$~eV by experiments above the atmosphere, the results being
well compatible with power laws\citelow{wiebel,pg}. Due to spallation during the
propagation process, the spectra of heavy elements are slightly flatter as
compared to light nuclei\citelow{pg,prop}, e.g.\ comparing protons
$\gamma_p=-2.71\pm0.02$ to iron nuclei $\gamma_{Fe}=-2.59\pm0.06$.

The regular component of the galactic magnetic field will cause particles with
charge $Z$ to describe helical trajectories with a Larmor radius $R_L=p/(Z e
B_0)= 1.08~\mbox{pc}\cdot E [\mbox{PeV}]/(Z\cdot B_0[\mu\mbox{G}])$, while the
random field component causes diffusive propagation. With increasing energy (or
momentum) it becomes more difficult to magnetically confine the particles to
the Galaxy.  Since $R_L\propto1/Z$ it is expected that leakage from the Galaxy
occurs for light elements (low $Z$) earlier as compared to heavy nuclei, i.e.\
protons leak first and subsequently all other elements start to escape from the
Galaxy.

\subsection{Structures in the Energy Spectrum}
Many possible origins for the knee are discussed in the
literature\citelow{origin,ecrsreview}. Most popular are assumptions of a
finite energy attained during the acceleration process and leakage from the
Galaxy as discussed. In both scenarios the energy spectra of elements exhibit a
cut-off at an energy proportional to the nuclear charge $Z$ and the knee in the
all-particle spectrum is caused by the cut-off of protons.

\ifnum1=2
Different ideas to explain the knee due to a finite energy attained during
acceleration have been discussed\citelow{berezhko,stanev,kobayakawa,sveshnikova}.
A special case of SNR acceleration is the single source model\citelow{wolfendale},
which proposes the knee being caused by CRs from a single SNR. 

As mentioned above, the pathlength of CRs in the Galaxy decreases as
$\Lambda\propto E^{-\delta}$. Such a decrease will ultimately lead to a
complete loss of the particles, with a rigidity dependent cut-off of the flux
for individual elements.  Many approaches have been undertaken to describe the
propagation process and a resulting knee.  Several diffusion models are
discussed\citelow{ptuskin,kalmykov,ogio,roulet,prop}.  During the propagation
phase, reacceleration of particles has been suggested at shock fronts in the
galactic wind\citelow{voelk}. Also a Leaky Box model\citelow{swordy} and an anomalous
diffusion model\citelow{lagutin} are proposed to explain the knee.

In the literature also other, more exotic ideas are mentioned. The acceleration
of particles in $\gamma$-ray bursts is described\citelow{plaga,wick,dar}.  Other
explanations for the knee are interactions of CRs with background
particles like massive neutrinos\citelow{dova,wigmans} or photo disintegration in
dense photon fields\citelow{tkaczyk,candia}.  A completely different reason for
the knee is the idea to transfer energy in nucleon-nucleon interactions into
particles, like gravitons\citelow{kazanas} or extremely high-energy
muons\citelow{petrukhin}, which are not observable (or not yet observed) in air
shower experiments.  However, some of these models already seem to be excluded
by present measurements\citelow{ecrsreview}.
\fi

All other elements follow subsequently and above a certain energy no more
particles are left.  On the other hand, the measured all-particle flux extends
up to $10^{20}$~eV, and the highest-energy particles are usually being
considered of extragalactic origin.  The Larmor radius of a proton with an
energy of $10^{20}$~eV in the galactic magnetic field is $R_L\approx36$~kpc,
comparable to the diameter of the Galaxy. This emphasizes that such high-energy
particles are of extragalactic origin.  The transition region from galactic to
extragalactic CRs is of particular interest, key features are the origin of the
second knee and the ankle.

Reviewing the properties of CRs accelerated in SNRs, Hillas finds that
a second (galactic) component is necessary to explain the observed flux at
energies above $10^{16}$~eV\citelow{hillasknee}.
Another possibility is a significant contribution of ultra-heavy elements
(heavier than iron) to the all-particle flux at energies around 400~PeV
\citelow{pg,prop}. In this approach the second knee is caused by the fall-off of
the heaviest elements with $Z$ up to 92. It is remarkable that the second knee
occurs at $E_{2nd}\approx92\times E_k$, the latter being the energy of the
first knee.
The dip seen in the spectrum between $10^{18}$ and $10^{19}$~eV, see
\fref{espec}, is proposed to be caused by electron-positron pair production of
CRs on cosmic microwave background photons\citelow{berezinskydip}
$p+\gamma_{3K}\rightarrow p+e^++e^-$.

\section{Measurement Techniques} \label{eass}
To clarify the situation and to distinguish between the different models,
measurements of the flux of individual elements, or at least groups of
elements, up to high energies are necessary.  Direct measurements above the
atmosphere on stratospheric balloons up to energies exceeding $10^{14}$~eV are
performed with various instruments like ATIC\citelow{atic}, CREAM\citelow{creamexp},
BESS\citelow{bess}, or TRACER\citelow{gahbauer}.  The presently largest experiment
with single-element resolution, TRACER, has an aperture of 5~m$^2$\,sr.  With
an exposure of 50~m$^2$\,sr\,d accumulated during a circumpolar flight in 2003
energy spectra could be measured up to about $5\cdot10^{14}$~eV for oxygen and
$8\cdot10^{13}$~eV for iron nuclei\citelow{tracer05}.

To extend the measurements to energies beyond the knee, at present, ground
based installations are the only possibility. With these experiments, secondary
products generated in the atmosphere are measured, the extensive air showers
(EAS).  Air showers were discovered in 1938 by W.
Kolh\"orster\citelow{kolhoersterschauer} and independently by P.
Auger\citelow{augerschauer}. Auger describes his work in a book\citelow{augerbuch}
translated by the director of this school M.M. Shapiro.

In todays experiments, the energy is basically derived from the number of
particles observed and the primary's mass is estimated by measurements of the
depth of the shower maximum (\sref{xmaxs}) or the electron-to-muon-ratio
(\sref{ems}).  Two types of experiments may be distinguished: installations
measuring the longitudinal development of showers in the atmosphere and
apparatus measuring the density (and energy) of secondary particles at ground
level.

An example for the latter is the KASCADE experiment\citelow{kascadenim}, covering
an area of $200\times200$~m$^2$. The basic idea is to measure the
electromagnetic component in an array of unshielded scintillation detectors and
the muons in scintillation counters shielded by a lead and iron absorber,
while the hadronic component is measured in a large calorimeter\citelow{kalonim}.
The total number of particles at observation level is obtained through the
measurement of particle densities and the integration of the lateral density
distribution\citelow{kascadelateral}.  The direction of air showers is
reconstructed through the measurement of the arrival time of the shower
particles in the individual detectors. 

The depth of the shower maximum is measured in two ways.  Light-integrating
\Cerenkov detectors like the BLANCA\citelow{blanca} or TUNKA\citelow{tunka}
experiments are in principle arrays of photomultiplier tubes with light
collection cones looking upwards in the night sky, measuring the lateral
distribution of \Cerenkov light at ground level. The depth of the shower
maximum and the shower energy is derived from these observations.  Imaging
telescopes as in the HiRes\citelow{hiresxmax} or AUGER\citelow{augerexp} experiments
observe an image of the shower on the sky through measurement of fluorescence
light, emitted by nitrogen molecules, which had been excited by air shower
particles.

\subsection{A Heitler Model for Air Showers}

The basic properties of EAS are illustrated using a Heitler model
\citelow{heitler}, expanding an approach by Matthews \citelow{matthewsheitler}.
The principle ideas of the model are emphasized by full EAS simulations
using the CORSIKA code \citelow{corsika} with the hadronic interaction models
FLUKA\citelow{flukacorsika} and QGSJET~01\citelow{qgsjet}. For the latter, a
modification with lower cross-sections has been used\citelow{wq}.
\footnote{Vertical showers with fixed energies between $10^5$ and
$3.16\cdot10^{10}$~GeV in steps of half a decade have been calculated.
Thresholds for photons, electrons, muons, and hadrons were chosen as
$E_\gamma>0.25$~MeV, $E_e>0.25$~MeV, $E_\mu>100$~MeV, and $E_h>100$~MeV.}

\subsubsection{Electromagnetic Cascades}
\begin{figure}[t] 
 \begin{minipage}[b]{0.49\textwidth}
  \psfig{file=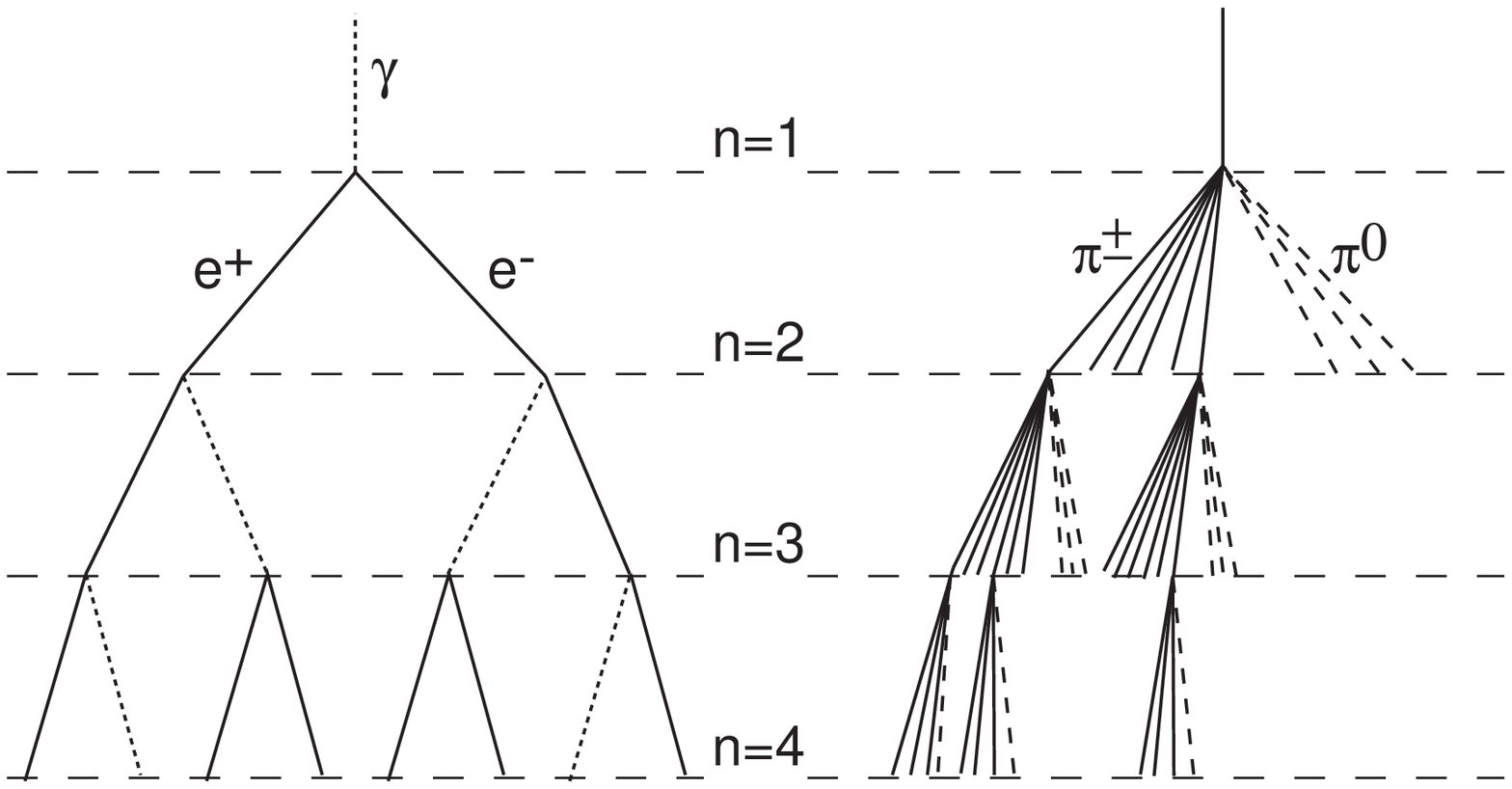,width=\textwidth}\vspace*{9mm}
 \end{minipage}
 \begin{minipage}[b]{0.49\textwidth}
  \psfig{file=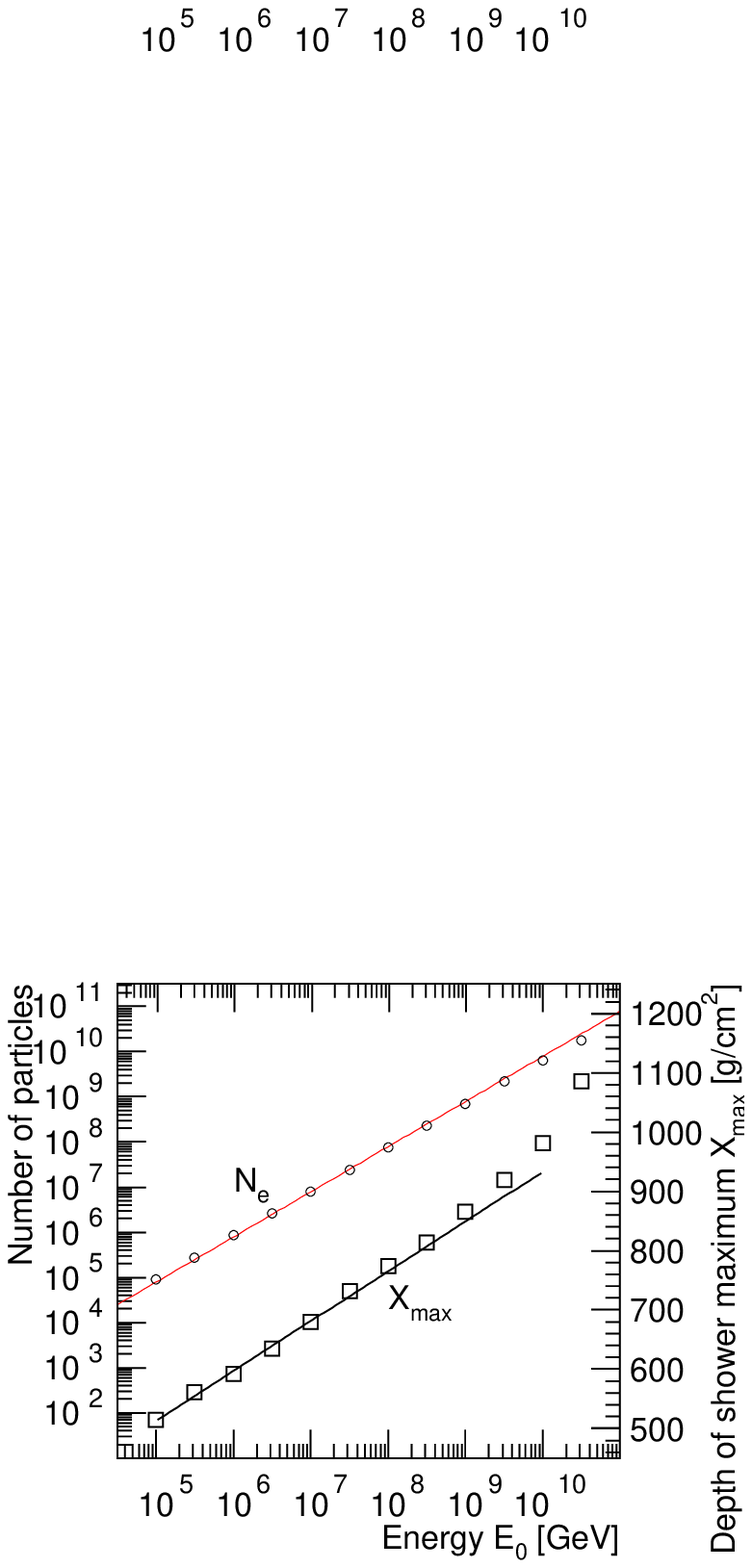,width=\textwidth}
 \end{minipage}\hspace*{\fill}
 \caption{\LLeft: Schematic view of an electromagnetic cascade (\lleft) and a
	  hadronic shower (\rright). Not all pion lines are shown
	  \Cite{matthewsheitler}.\label{heitlermod}
          \RRight: Number of electrons at shower maximum and depth of the
	  shower maximum as function of photon energy. The lines are according
	  to \eref{xmaxgamma} and \eref{nemax}.  \label{emschauer}}	  
\end{figure}

A simple approximation of an electromagnetic cascade is shown schematically in
\fref{heitlermod}. A primary photon generates an $e^+e^-$ pair. An electron
radiates a single photon after traveling one splitting length $d=X_0\ln2$,
where $X_0$ is the radiation length ($X_0^{air}=36.66$~\gcm2). An electron
looses on average half of its energy through radiation over the distance $d$.
After traveling the same distance a photon splits into an $e^+e^-$ pair. In
either instance, the energy of a particle is assumed to be equally divided
between two outgoing particles.  After $n$ splitting lengths, at a distance
$x=n X_0\ln2$, the total shower size (electrons and photons) is
$N=2^n=\exp(x/X_0)$ and the initial energy $E_0$ is distributed over $N$
particles.  The splitting continues until the energy per particle $E_0/N$ is
too low for pair production or bremsstrahlung.  Heitler takes this energy to be
the critical energy ($E_c^e=85$~MeV in air), at which ionization losses and
radiative losses are equal.

A shower initiated by a primary photon reaches its maximum size $N_{max}$ when
all particles have the energy $E_c^e$, which means $E_0=E_c^e N_{max}$.  The
penetration depth \Xmax at which the shower reaches its maximum is obtained by
determining the number $n_c$ of splitting lengths, required to reduce the
energy per particle to $E_c^e$.  Since $N_{max}=2^{n_c}$, the number of
splitting lengths is $n_c=\ln(E_0/E_c^e)/\ln2$, giving $N_{max}=E_0/E_c^e$ and
\begin{equation} \label{xmaxgamma}
  X_{max}^\gamma=n_cX_0\ln2 = X_0\ln(E_0/E_c^e) .
\end{equation}
The elongation rate $\Lambda$ specifies the increase of \Xmax with energy $E_0$
and is defined as $\Lambda\equiv \mbox{d}X_{max}/\mbox{d}\,\lg E_0$.  Using
\eref{xmaxgamma} gives $\Lambda^\gamma=\ln10 \,X_0=84.4$~\gcm2 per decade of
primary energy for electromagnetic showers in air.  Thus,
$X_{max}=597~\mbox{\gcm2}+84~\mbox{\gcm2}\,\lg(E_0/\mbox{PeV})$ is expected.
This prediction agrees well with results of full simulations as can be inferred
from \fref{emschauer}. \footnote{The deviations at high energies are due to the
Landau Pomeranchuk Migdal effect\citelow{pomeranchuk,migdal}.}

The simple model describes quite well the position of the maximum of
electromagnetic cascades when compared to EAS simulations and to
measurements at accelerators\citelow{matthewsheitler,kalocern}.  However, the
model overestimates the actual ratio of electrons to photons. It predicts that
after a few generations the electron size approaches
$N_e\approx\frac{2}{3}N_{max}$. This is much too large for several reasons,
mainly that multiple photons are often radiated during bremsstrahlung and many
electrons and positrons range out in the air. 

To extract the number of electrons $N_e^{max}$ at shower maximum from Heitler's
total size $N_{max}$, a simple correction 
\begin{equation}\label{nemax} 
 N_e^{max}=\frac{E_0}{g E_c^e} 
\end{equation} 
is adopted, with a constant value $g$.  When the estimated electron number is
compared to measurements, the factor $g$ has to be fine tuned. It depends on
properties of the detectors used like the energy threshold and the efficiency
to detect photons and electrons (or positrons). Comparisons with results at
accelerators indicate values between $g=10$\citelow{matthewsheitler} and $g=20$
\citelow{kalocern}.  Results of a full EAS simulation are depicted in
\fref{emschauer}. For electromagnetic showers the number of electrons turns out
to be almost exactly linearly proportional to the shower energy as expected
from \eref{nemax}. A fit yields $N_e\propto E_0^{0.97}$ and a correction factor
$g\approx13$ is obtained, compatible with the accelerator based results.
With this value the number of electrons at shower maximum is according to
\eref{nemax} $N_e^{max}=9.0\cdot10^5\cdot E_0$/PeV.

\subsubsection{Hadronic Showers}

Hadron induced showers are modeled using a similar approach, for a figurative
sketch, see \fref{heitlermod}. The atmosphere is divided in layers of fixed
thickness $\lambda_i\ln2$, where \li is the interaction length of strongly
interacting particles.  An energy around 100~GeV is a typical energy for pions
in air showers and for a simple approach a constant value $\lambda_i=120$~\gcm2
is adopted. 
Hadrons interact after traversing one layer, producing $N_{ch}$ charged pions
and $\frac{1}{2}N_{ch}$ neutral pions. The latter decay promptly to photons,
initiating electromagnetic cascades. Charged pions continue through another
layer and interact. The process continues until the charged pions fall bellow
the critical energy $E_c^\pi$, where they are all assumed to decay, yielding
muons.

The multiplicity of charged particles produced in hadron interactions increases
very slowly with laboratory energy $\propto E^{0.2}$ in $pp$ and $p\bar p$ data
\citelow{pdg04}. The multiplicity in $\pi$-$^{14}$N collisions increases as
$N_{ch}\approx5$, 11, and 27 at 10, 100, and $10^4$~GeV,
respectively\citelow{wq}.  A constant value $N_{ch}=10$ is adopted in the
following for the number of charged particles produced in pion-air
interactions, again corresponding to an energy of about 100~GeV.

The second parameter is the energy $E_c^\pi$ at which further particle
production by $\pi^\pm$ ceases.  $E_c^\pi$ may be defined as the energy at
which the probability for decay and hadronic interaction equalize.  Following
\rref{matthewsheitler} a constant critical pion energy $E_c^\pi=20$~GeV is
adopted in the following.

If we consider a proton with $E_0$ entering the
atmosphere, we have after $n$ interactions $N_\pi=(N_{ch})^n$ charged pions.
Assuming equal division of energy during particle production, these pions carry
a total energy of $(2/3)^n E_0$. The remainder of the energy goes into
electromagnetic showers from $\pi^0$ decays. Hence, the energy per charged pion
is $E_\pi=E_0/(\frac{3}{2}N_{ch})^n$.  After a certain number $n_c$ of
generations, $E_\pi$ becomes less than $E_c^\pi$.
The number of interactions needed to reach $E_\pi=E_c^\pi$ is
\begin{equation} \label{nceq}
 n_c= \frac{\ln E_0/E_c^\pi}{\ln\frac{3}{2}N_{ch}}
    = 0.85\,\lg\left(\frac{E_0}{E_c^\pi}\right) .
\end{equation}

\subsubsection{Number of Muons}

The number of muons is obtained, assuming that all pions decay, using
$N_\mu=N_\pi=(N_{ch})^{n_c}$. Their energy dependence is derived applying
\eref{nceq}
\begin{equation} \label{betaeq}
 \ln N_\mu=n_c\ln N_{ch}=\beta\ln\left(\frac{E_0}{E_c^\pi}\right) , 
 \quad\mbox{with}\quad \beta=\frac{\ln N_{ch}}{\ln\frac{3}{2}N_{ch}}\approx0.85
\end{equation}
for $N_{ch}=10$. It should be noted that although $N_{ch}$ changes (slowly) as
the shower develops, $\beta$ depends only logarithmically on this value.

So far, an important aspect of hadronic interactions has been neglected. In an
interaction only a fraction of the energy is available for secondary particle
production, usually characterized by the the inelasticity $\kappa$.  Taking
this effect into account, in an interaction initiated by a particle with energy
$E$, the energy $(1-\kappa)E$ is taken away by a single leading particle,
$\frac{2}{3}\kappa E$ is used to produce $N_{ch}$ charged pions, and
$\frac{1}{3}\kappa E$ goes via neutral pions into the electromagnetic
component.  Including inelasticity in the Heitler model changes the parameter
$\beta$ in \eref{betaeq} to\citelow{matthewsheitler}
\begin{equation} \label{betakappa}
 \beta=\frac{\ln[1+N_{ch}]}{\ln\left[(1+N_{ch})/(1-\frac{1}{3}\kappa)\right]}
      \approx1-\frac{\kappa}{3\ln(N_{ch})}=1-0.14\kappa .
\end{equation}
The elasticity for the most energetic meson in pion-air interactions yields
$1-\kappa$ between 0.26 and 0.32\citelow{wq}, resulting in $\beta=0.90$.

To expand the simple approach from primary protons to nuclei, the superposition
model is used. A nucleus with atomic mass number $A$ and energy $E_0$ is taken
to be $A$ individual single nucleons, each with energy $E_0/A$, and each acting
independently. The resulting EAS is treated as the sum of $A$ individual proton
induced showers, all starting at the same point. The observable shower features
are obtained by substituting the lower primary energy into the expressions
derived for proton showers and summing $A$ such showers.  Applying this to the
number of muons yields $N_\mu=A(E_0/(A E_c^\pi))^\beta$.  The number of muons
in showers induced by nuclei with mass number $A$ and energy $E_0$ is then
\begin{equation} \label{nmueq}
 N_\mu = \left(\frac{E_0}{E_c^\pi}\right)^\beta A^{1-\beta}
         \approx1.69\cdot10^4 \cdot
	 A^{0.10} \left(\frac{E_0}{1~\mbox{PeV}}\right)^{0.90} .
\end{equation}
Two important features follow from \eref{nmueq}: the number of muons increases
as function of energy slightly less than exactly linear and $N_\mu$ increases
as function of the mass of the primary particle as $\propto A^{0.1}$.
Accordingly, iron induced showers contain about 1.5 times as many muons as
proton showers with  the same energy.  This results from the less-than-linear
growth of the number of muons with energy -- $\beta<1$ in \eref{nmueq}.  The
lower energy nucleons which initiate the shower generate fewer interaction
generations, and consequently, loose less energy to the electromagnetic
component. 

\begin{figure}[t] 
 \begin{minipage}[t]{0.49\textwidth}
  \psfig{file=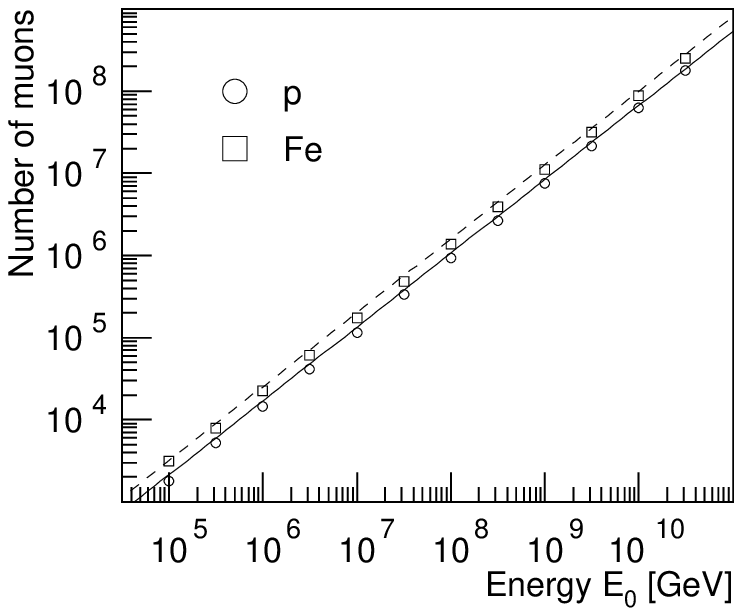,width=\textwidth}%
 \end{minipage}\hspace*{\fill}
 \begin{minipage}[t]{0.49\textwidth}
  \psfig{file=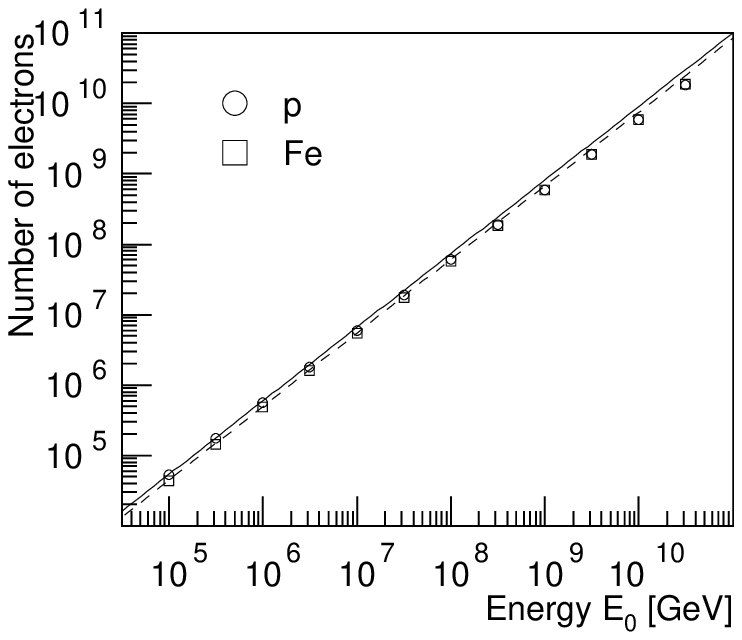,width=\textwidth}
 \end{minipage}
 \caption{Number of muons (\lleft) and number of electrons (\rright) at shower
	  maximum as function of energy for primary protons and iron nuclei
	  according to CORSIKA simulations (symbols). The lines are predictions
	  according to \eref{nmueq} and \eref{neeq}, respectively, for protons
	  (\line) and iron (\dashed) nuclei.}
 \label{nenm}	  
\end{figure}

The number of muons at shower maximum as function of energy is shown in
\fref{nenm} as obtained from full simulations.  The lines indicate predictions
according to \eref{nmueq}, being well in agreement with the simulations.

\subsubsection{Number of Electrons}
Conservation of energy implies that the primary energy is split into
electromagnetic and hadronic parts $E_0=E_{em}+E_h$. The number of
electrons is estimated using this relation. The hadronic energy appears in the
simple approach in the muon component as $E_h=N_\mu E_c^\pi$ and the energy
fraction for the electromagnetic component is, using \eref{nmueq}
\begin{equation} \label{emfracteq}
 \frac{E_{em}}{E_0}= \frac{E_0-N_\mu E_c^\pi}{E_0}
                   = 1-\left(\frac{E_0}{A\,E_c^\pi}\right)^{\beta-1} .
\end{equation}
The electromagnetic fraction is 57\% at $E_0=10^{14}$~eV, increasing to 79\%
at $10^{17}$~eV for proton induced showers. For iron induced showers the
fraction rises from 38\% to 68\%.

\Eref{emfracteq} can be approximated by a power law 
\begin{equation}
 \frac{E_{em}}{E_0}\approx a \left(\frac{E_0}{A\,E_c^\pi}\right)^b .
\end{equation}
Series expansion near $x_0=E_0/E_c^\pi=10^5$ yields the number of electrons at
shower maximum as function of energy
\begin{equation} \label{ne}
 N_e = \frac{E_{em}}{g E_c^e}
       \approx \frac{a}{g E_c^e} \left(A\,E_c^\pi\right)^{-b} E_0^{1+b} 
\end{equation}
with $b=(1-\beta)/(x_0^{1-\beta}-1) \approx 0.046$ and
$a=(1-x_0^{\beta-1})/(x_0^b)\approx 0.40$.  Hence, $\alpha=1+b\approx1.046$ is
obtained, which leads, using $g=13$ to
\begin{equation} \label{neeq}
   N_e \approx 5.95\cdot10^5\cdot A^{-0.046}
       \left(\frac{E_0}{1~\mbox{PeV}}\right)^{1.046}	.
\end{equation}
This implies that the number of electrons grows as function of energy slightly
faster than exactly linear. The electron number decreases with increasing mass
number, an iron induced shower is expected to contain about 83\% of the
electromagnetic energy of a proton shower with the same energy.  It should be
emphasized that the model does not take into account absorption in the
atmosphere, thus, the number of electrons obtained is valid at shower maximum.

The number of electrons at shower maximum according to full simulations is
shown as function of energy in \fref{nenm}.
The results are compared to predictions according to \eref{neeq} for
proton and iron induced showers indicated by the lines.
It can be seen that the simple model reproduces quite well the results of the
full simulations.

\subsubsection{Depth of the Shower Maximum} \label{xmaxs}

The atmospheric depth at which the electromagnetic shower component reaches its
maximum is called \Xmax.  In hadronic interactions $\kappa/3$ of the available
energy goes into the electromagnetic component via $\pi^0$-decays, see
\fref{heitlermod}.  For a simple estimate only the first generation of
electromagnetic showers is used.  This approach will certainly underestimate
the value of \Xmax since it neglects the following subshowers.

The first interaction occurs at an atmospheric depth $X_1=\lambda_i\ln2$, where
$\lambda_i$ is the interaction length of a primary proton $\lambda_i^{p-air}$.
The latter can be approximated around 1~PeV by the relation
\begin{equation} \label{lambdaeq}
 \lambda_i^{p-air}=\xi+\zeta\,\lg \frac{E_0}{1~\mbox{PeV}}
\end{equation}
with $\xi=68.55$~\gcm2 and $\zeta=-4.88$~\gcm2.

In the first interaction $\frac{1}{2}N_{ch}$ neutral pions are produced,
yielding $N_{ch}$ photons. Each photon initiates an electromagnetic cascade
with the energy $\kappa E_0/(3N_{ch})$, developing in parallel with the others.
The average multiplicity of charged particles produced in pion-nitrogen
interactions\citelow{wq} can be parameterized for energies around 1~PeV as
\begin{equation} \label{ncheq}
 N_{ch}=N_0 \left(\frac{E_0}{1~\mbox{PeV}}\right)^\eta 
\end{equation}
with $N_0=55.2$ and $\eta=0.13$.

The depth of the shower maximum is obtained as in \eref{xmaxgamma} for an
electromagnetic shower with an energy $\kappa E_0/(3N_{ch})$, starting after the
first interaction at a depth $X_1$,
$X_{max}^p=\lambda_i^{p-air} \ln2 + X_0\ln(\kappa E_0/(3N_{ch}E_c^e))$.
Using \eref{lambdaeq} and \eref{ncheq}, the expression
\begin{equation} \label{xmaxp}
  X_{max}^p= \xi\ln2 
          - X_0\ln\left(\frac{3N_0}{\kappa}\frac{E_c^e}{\mbox{PeV}}\right)
  + \left( X_0\ln10 -\eta X_0\ln10+\zeta\ln2 \right)\lg \frac{E_0}{\mbox{PeV}}
\end{equation}
is obtained.  The elongation rate for protons is determined by the elongation
rate for electromagnetic showers $\Lambda^\gamma=X_0\ln10$ and in addition by
terms which take into account the growing multiplicity of secondary particles,
as well as the decreasing interaction length as function of energy.  Taking the
numerical parameters as described, the elongation rate
$\Lambda^p=(84.4-11.0-3.4)~\mbox{\gcm2}=70.0~\mbox{\gcm2}$ is obtained, and one
realizes that the effect of growing multiplicity dominates the effect of a
decreasing interaction length by about a factor three. Evaluating also the
constant term in \eref{xmaxp} yields
$X_{max}^p=442.9~\mbox{\gcm2}+70.0~\mbox{\gcm2}\,\lg (E_0/\mbox{PeV})$.

\begin{figure}[t] 
 \begin{minipage}[b]{0.49\textwidth}
  \psfig{file=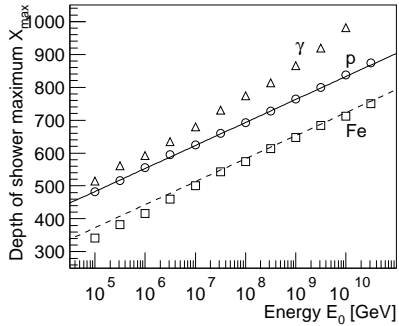,width=\textwidth}
 \end{minipage}
 \begin{minipage}[b]{0.49\textwidth}
 \caption{Average depth of the shower maximum for primary photons,
	  protons, and iron nuclei according to CORSIKA simulations.  The lines
	  indicate predictions according to \eref{xmaxp} (\dashed) and the same
	  function shifted up by 110~\gcm2 (\line).}
 \label{xmaxheitler}	  
 \end{minipage}
\end{figure}

When compared to results of full simulations, the calculated values for
$X_{max}^p$ are about 110~\gcm2 or almost $2\lambda_i^{p-air}$ smaller than the
results of full calculations depicted in \fref{xmaxheitler} (dashed line).
Presumably this is a consequence of neglecting the contributions of following
generations of $\pi^0$ production.  However, the predicted elongation rate
agrees extremely well with the value obtained from the CORSIKA simulations at
1~PeV $\Lambda^p=69.9\pm0.3$~\gcm2 per decade. The solid line represents
\eref{xmaxp} shifted upwards by 110~\gcm2 and agrees well with the proton
simulations.

To expand the simple approach from primary protons to nuclei with mass number
$A$, the superposition model is used and in \eref{xmaxp} the energy $E_0$ is
substituted by $E_0/A$. This yields $X_{max}^A=X_{max}^p - X_0 \ln A$,
predicting that the maximum for iron induced showers should be about 150~\gcm2
higher up in the atmosphere.  In the full simulations, the difference is
slightly smaller as can be inferred from the figure.

\subsubsection{Energy and Mass of the Primary Particle} \label{ems}

In EAS experiments the reconstructed number of electrons and muons are
often presented in the $\lg N_\mu$-$\lg N_e$ plane in order to estimate the
energy and mass of the shower inducing particles. As an application of the
simple Heitler Model, lines of constant mass and energy in the $N_\mu$-$N_e$
plane are derived in the following to illustrate the method utilized in the
experiments.

To deduce lines of constant mass, \eref{ne} is transformed to obtain $E_0$,
which, in turn is introduced into \eref{nmueq}. This yields the number of muons
as function of the number of electrons at shower maximum
\begin{equation} \label{aline}
 \left. N_\mu \right|_{A}
     = (E_c^\pi)^{-\delta} \left(\frac{g E_c^e}{a}\right)^\delta 
         A^{1-\delta} N_e^\delta
       \approx 0.18 \cdot A^{0.14} N_e^{0.86}	 
\end{equation}
with the exponent $\delta = \beta/(1+b) \approx 0.86$.  In a similar way, lines
of constant energy are derived. $A$ is taken from \eref{ne} and put in
\eref{nmueq}, which leads to
\begin{equation} \label{eline}
 \left. N_\mu \right|_{E_0} =
      \frac{1}{E_c^\pi} \left(\frac{g E_c^e}{a}\right)^\varepsilon
        E_0^{\beta+\varepsilon(b-1)} N_e^\varepsilon
   \approx 5.77\cdot10^{16} \left(\frac{E_0}{1~\mbox{PeV}}\right)^{2.97}
      N_e^{-2.17}	
\end{equation}
with an exponent $\varepsilon=-(1-\beta)/b \approx -2.17$.

\begin{figure}[t] 
 \begin{minipage}[t]{0.49\textwidth}
  \psfig{file=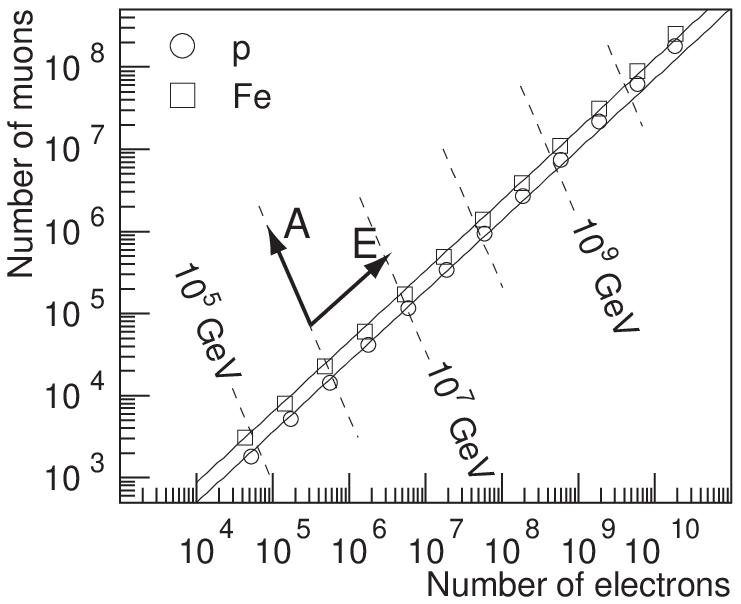,width=\textwidth}%
 \end{minipage}\hspace*{\fill}
 \begin{minipage}[t]{0.49\textwidth}
  \psfig{file=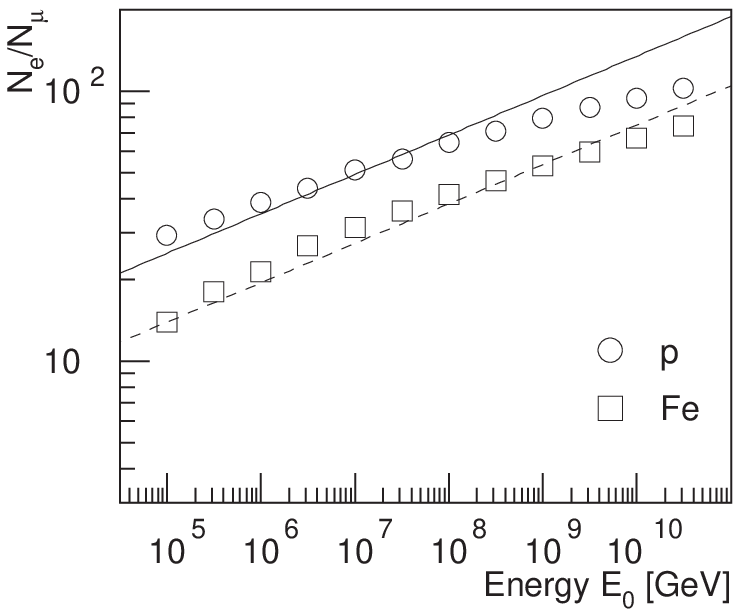,width=\textwidth}
 \end{minipage}
 \caption{\LLeft: Average number of muons versus number of electrons at shower
	  maximum for primary protons and iron nuclei.  The data points are
	  results of full simulations. The solid lines represent \eref{aline}
	  for protons and iron nuclei, the dashed lines are equal energy lines
	  according to \eref{eline}.
	  \RRight: Ratio of electron to muon number $N_e/N_\mu$ at shower
	  maximum as function of energy for primary protons and iron nuclei.
	  The data points are results of full simulations. The lines indicate
	  \eref{emratioeq} for the two primaries.}
 \label{emplane}	  
\end{figure}

The constant-mass lines for protons and iron nuclei are shown in \fref{emplane}
together with equal-energy lines for energies from $10^5$ to $10^{10}$~GeV.
These sets of lines form a coordinate system for energy and mass in the
$N_\mu$-$N_e$ plane. The axis are non-perpendicular to each other.  In the
figure also results of full CORSIKA simulations for proton and iron induced
showers are shown for fixed energies from $10^5$ to $3.16\cdot10^{10}$~GeV in
steps of half a decade.  Taking the simplicity of the model into account the
predicted lines agree quite well with the full simulations and they give a good
illustration of the physics in the $N_\mu$-$N_e$ plane.

Dividing \eref{ne} by \eref{nmueq} yields the electron-to-muon ratio at shower
maximum
\begin{equation} \label{emratioeq}
 \frac{N_e}{N_\mu} = \frac{a}{g E_c^e} (E_c^\pi)^{\beta-b} 
                     \left(\frac{E_0}{A}\right)^{1+b-\beta}
   \approx 35.1 \cdot \left(\frac{E_0}{1~\mbox{PeV}\cdot A}\right)^{0.15} .
\end{equation}
It depends on the energy per nucleon $E_0/A$ of the primary particle.  This is
the reason why the ratio $N_e/N_\mu$ is frequently used in EAS experiments to
estimate the mass of the primary particle.  If the energy is derived from
another observable, the mass can be inferred.  Predictions according to
\eref{emratioeq} are compared to results of full simulations for proton and
iron induced showers in \fref{emplane}. The simple model predicts the
calculated ratio quite well.  The CORSIKA simulations exhibit almost a power
law behavior, however, at high energies some flattening with respect to the
predicted slope is visible. 

\section{Experimental Results} \label{ress}

The all-particle energy spectra obtained by many experiments are compiled in
\fref{espec}. Shown are results from direct measurements above the atmosphere
as well as from various air shower experiments. The individual measurements
agree within a factor of two or three in the flux and a similar shape can be
recognized for all experiments with a knee at energies of about 4~PeV.  Typical
values for the systematic uncertainties of the absolute energy scale for air
shower experiments are about 15 to 20\%.  Renormalizing the energy scales of
the individual experiments to match the all-particle spectrum obtained by
direct measurements in the energy region up to almost a PeV requires correction
factors of the order of $\pm10$\%\citelow{pg}.  Indicating that the all-particle
spectrum seems to be well determined.

Due to the large fluctuations in air showers it is not possible to derive
energy spectra for individual elements from air shower data.  Therefore,
frequently the mean mass of CRs is investigated.  An often-used quantity to
characterize the composition is the mean logarithmic mass, defined as $\lnA=
\sum_i r_i \ln A_i$, $r_i$ being the relative fraction of nuclei of mass $A_i$.
Investigating the ratio of the number of electrons and muons at ground level
and the average depth of the shower maximum an increase of the mean logarithmic
mass in the energy range around the knee could be observed by many experiments
\citelow{aspenreview}. Such an increase is expected from consecutive cut-offs of
the energy spectra of individual elements, starting with protons.

\begin{figure}[t]\centering
  \psfig{file=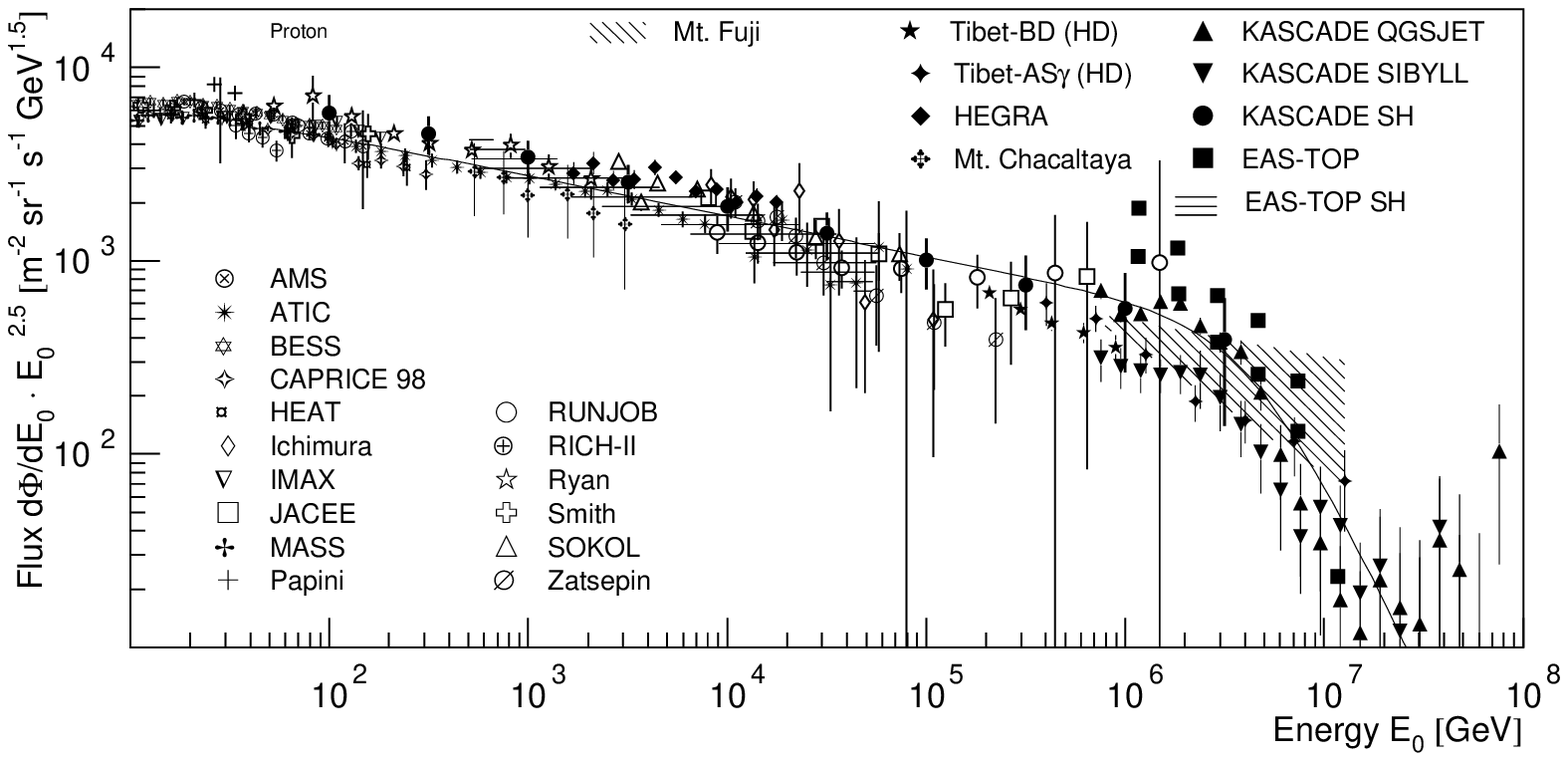,width=0.85\textwidth}
  \psfig{file=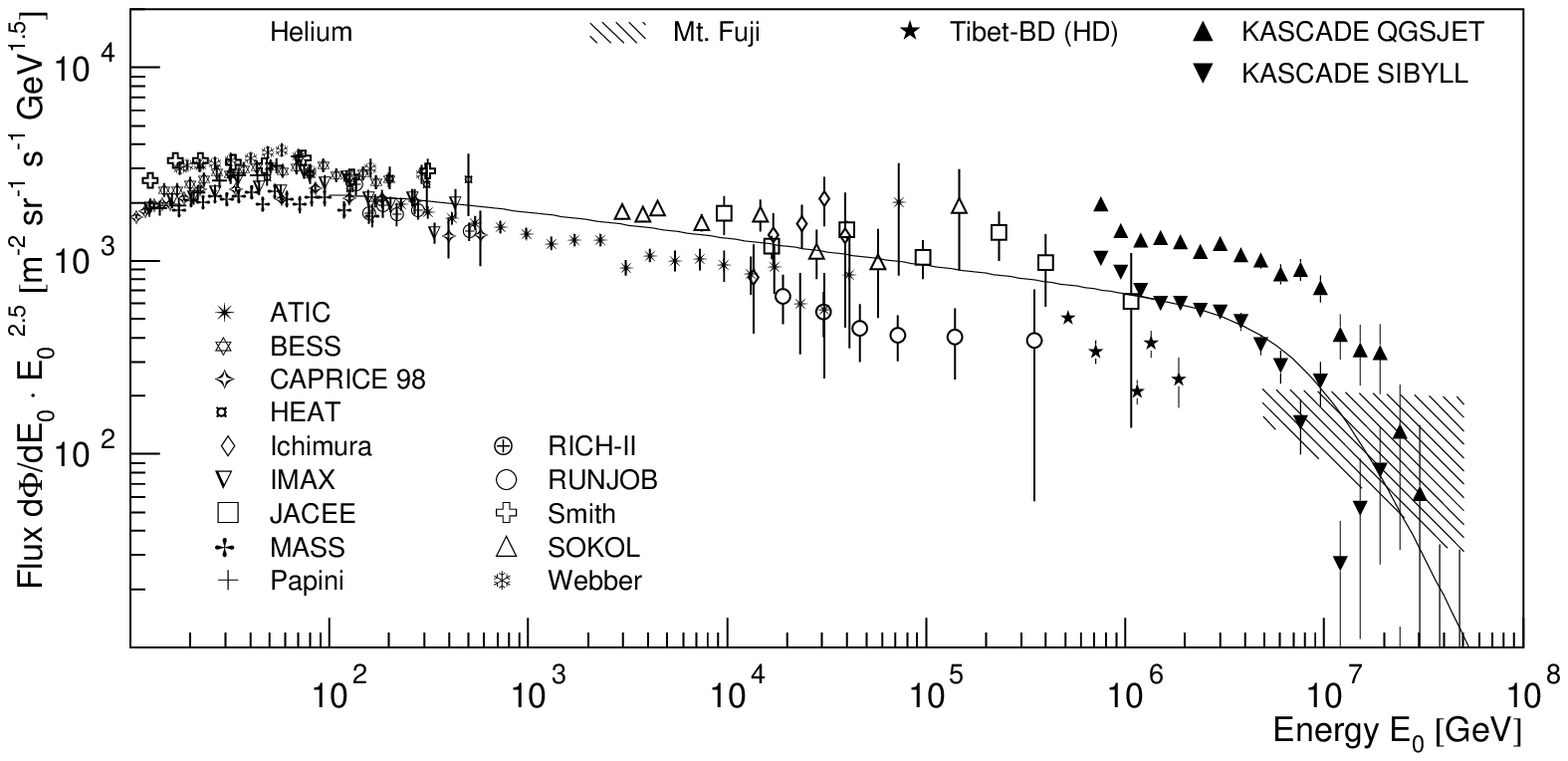,width=0.85\textwidth}
  \psfig{file=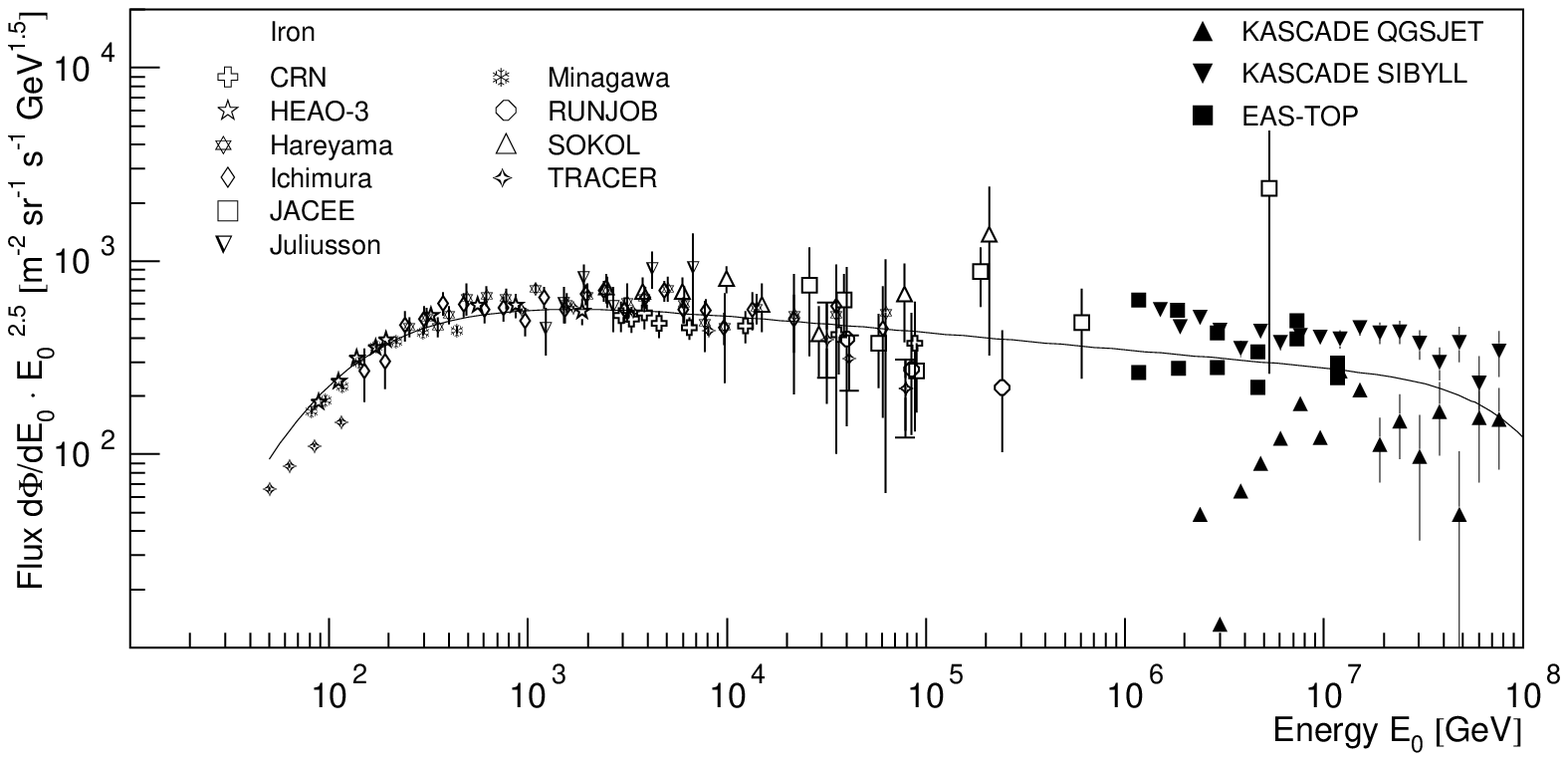,width=0.85\textwidth}
  \caption{Energy spectra for primary protons, helium, and iron nuclei
	   from direct and indirect measurements for references
	   see\Cite{aspenreview}. The lines indicate spectra according to the
	   poly-gonato model\Cite{pg}. \label{elementspec}}
\end{figure}

A significant step forward in understanding the origin of CRs are measurements
of energy spectra for individual elements or at least groups of elements.  Up
to about a PeV direct measurements have been performed with instruments above
the atmosphere. As examples, results for primary protons, helium, and iron
nuclei are compiled in \fref{elementspec}.  Recently, also indirect
measurements of elemental groups became possible.

\begin{figure}[t]\centering
 \begin{minipage}{0.49\textwidth}
  \psfig{file=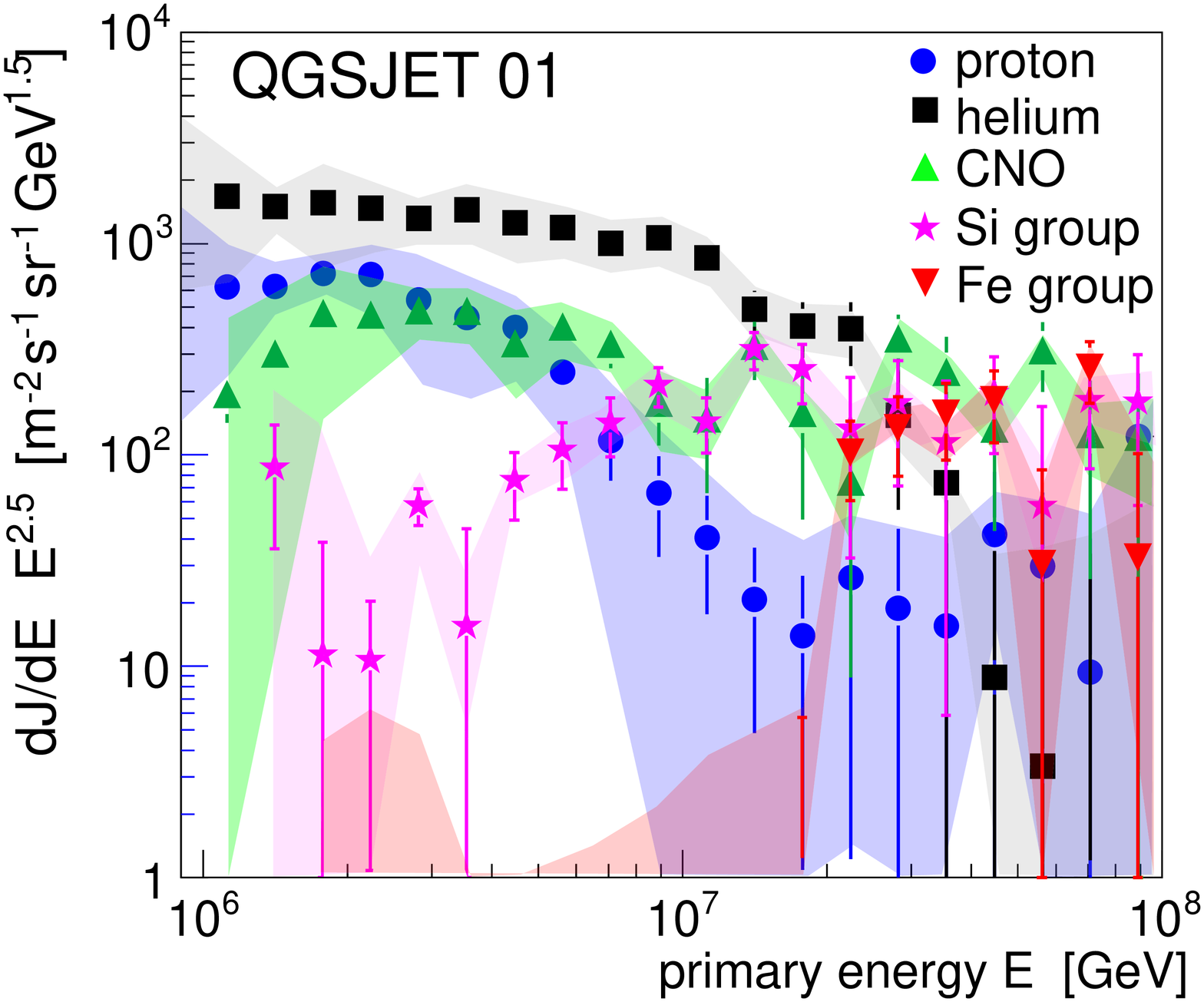,width=\textwidth}%
 \end{minipage} 
 \begin{minipage}{0.49\textwidth}
  \psfig{file=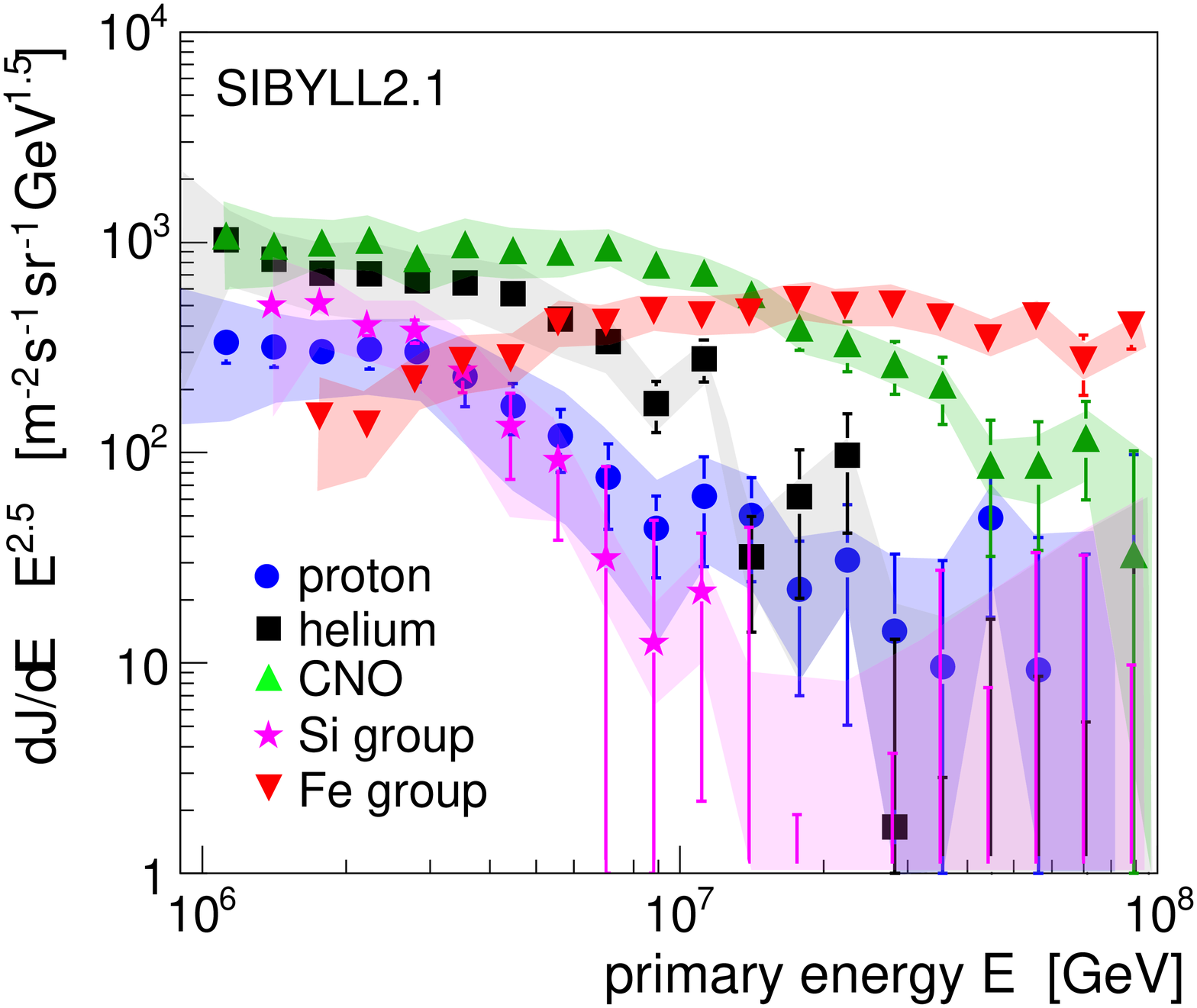,width=\textwidth}%
 \end{minipage} 
  \caption{Energy spectra for elemental groups as obtained by the KASCADE
	   experiment, using two different models (QGSJET~01 and SIBYLL~2.1) to
	   interpret hadronic interactions in the
	   atmosphere\Cite{ulrichapp}.\label{kascade}}
\end{figure}

With the KASCADE experiment, the problem of composition has been approached in
various ways\citelow{jrhtaup05}.  An advanced analysis is founded on the
measurement of the electromagnetic and muonic shower
components\citelow{ulrichapp}.  It is based on the deconvolution of a
two-dimensional electron muon number distribution.  Unfolding is performed
using two hadronic interaction models (QGSJET\,01 and SIBYLL\,2.1) to interpret
the data. The spectra obtained for five elemental groups are displayed in
\fref{kascade}. They exhibit sequential cut-offs in the flux for the light
elements.  For both models a depression is visible for protons around 3 to
4~PeV and at higher energies for helium nuclei.  The systematic differences in
flux for the spectra derived with QGSJET and SIBYLL amount to a factor of about
two to three.  The silicon and iron groups show a rather unexpected behavior
for both models.  The increase of the flux for both groups (QGSJET) and the
early cut-off for the silicon group (SIBYLL) is not compatible with
contemporary astrophysical models.  The discrepancies are attributed to the
fact that none of the air shower models is able to describe the observed data
set in the whole energy range consistently \citelow{ulrichapp}. 

The KASCADE results are compared to results of other experiments in
\fref{elementspec}.
EAS-TOP derived spectra from the simultaneous observation of the
electromagnetic and muonic components.
HEGRA used an imaging \Cerenkov telescope system to derive the primary proton
flux\citelow{hegrap}.
The primary proton flux has been derived from measurements of the flux of
unaccompanied hadrons at ground level with the EAS-TOP and KASCADE
experiments\citelow{eastopsh,kascadesh}.  
Spectra for protons and helium nuclei are obtained from emulsion chambers
exposed at Mts.\ Fuji and Kanbala\citelow{mtfujip}.
The Tibet group performs measurements with a burst detector as well as with
emulsion chambers and an air shower array\citelow{tibetbdp,tibetasg03}.

Considering the energy range above 10~GeV, at least a qualitative picture of
the energy spectra for individual mass groups emerges: the spectra seem to be
compatible with power laws with a cut-off at high energies.  The spectra
according to the poly gonato model \citelow{pg} are indicated in the figures as
lines. It can be recognized that the measured values are compatible with
cut-offs at energies proportional to the nuclear charge
$\hat{E}_Z=Z\cdot4.5$~PeV.  The lines in \fref{espec} indicate spectra for the
same model.  Summing up the flux of all elements, the all-particle flux is
compatible with the flux derived from air shower experiments in the knee
region. Above $10^8$~GeV the flux of galactic CRs is not sufficient to account
for the observed all-particle spectrum, and an additional, presumably
extragalactic component is required. 

Energy spectra have been reconstructed with KASCADE data up to energies of
100~PeV. At these energies statistical errors start to dominate the overall
error. To improve this situation, the experiment has been enlarged.  Covering
an area of 0.5~km$^2$, 37 detector stations, containing 10~m$^2$ of plastic
scintillators each, have been installed to extend the original KASCADE set-up
\citelow{grande}.  Regular measurements with this new array and the original
KASCADE detectors, forming the KASCADE-Grande experiment, are performed since
summer 2003\citelow{chiavassapune}.  The objective is to reconstruct energy
spectra for groups of elements up to $10^{18}$~eV\citelow{haungsaspen}, covering
the energy region of the second knee, where the galactic cosmic ray spectrum is
expected to end\citelow{aspenphen}.
First analyses extend the lateral distributions of electrons and muons up to
600~m\citelow{glasstetterpune,vanburenpune}.  Based on one year of measurements,
already energies close to $10^{18}$~eV are reached. It is planned to conduct an
unfolding analysis, similar to the one described above, and reveal the energy
spectra for groups of elements up to $10^{18}$~eV.

A more detailed discussion of experimental results may be found
elsewhere\citelow{chicagoknee,aspenreview,ecrsreview,jrherice04}.

\section{Conclusion and Outlook}
In the last decade the understanding of the origin of high-energy CRs
has advanced significantly.  In particular, the KASCADE experiment has shown
that the origin of the knee in the all-particle energy spectrum is due to a
cut-off of the light elements. A corresponding increase of the mean mass as
function of energy in the knee region is observed by many experiments.  Such a
behavior is expected from astrophysical models, explaining the knee due to a
finite energy reached in the acceleration process and due to leakage from the
Galaxy.  However, it has also evolved that the astrophysical interpretation of
air shower data, at present, is limited by the understanding of high-energy
hadronic interactions in the atmosphere. Experiments like KASCADE have reached
the sensitivity to improve interaction models and corresponding analyses are
under way, e.g.\ \citelow{isvhecri04kascadewq,kascadewqpune,jenspune}. A big step
forward is the observation of TeV-$\gamma$-rays from SNRs with the expected
spectral index $\gamma\approx-2.1$, thus giving an important hint to the
sources of hadronic CRs.

In the next years the KASCADE-Grande experiment and the Ice Cube/Ice Top
experiment at the south pole\citelow{icetop} will measure CRs in the
energy region of the second knee and will provide information on the mass
composition in this region, where the galactic CR component is expected to end.
Balloon borne experiments like ATIC, CREAM, or TRACER will improve the
knowledge about CR propagation, by extending the energy spectra of individual
elements to energies approaching the knee.

\section*{Acknowledgment} 
It was a pleasure to participate in an inspiring school and to experience the
great hospitality in Erice.  I would like to thank Jim Matthews for our
exchange about the Heitler model and acknowledge the stimulating scientific
discussions with my colleagues from the KASCADE-Grande, TRACER, and AUGER
experiments.


\end{document}